# Simplify Power Flow Calculation Using Terminal Circuit and PMU Measurements

Renchang Dai[1], *Senior Member, IEEE*, Xiang Zhang[1], *Student Member, IEEE*, Junjie Shi[2], Guangyi Liu[1], *Senior Member, IEEE*, Chen Yuan[1], *Member, IEEE,* Zhiwei Wang[1], *Senior Member, IEEE*

[1] Global Energy Interconnection Research Institute, San Jose, CA 95134, USA

[2] State Grid Sichuan Electric Power Company, Sichuan, China

***Abstract*—Power flow calculation methods have been developed in decades using power injections and Newton-Raphson method. The nonlinear characteristics of the power flow to the bus voltage require Jacobian matrix reformation and refactorization in each iteration. Power network is composed by resistors, reactors, and capacitors which is a linear circuit when investigating the node voltages with the node current injections. To take the advantage of the linearity, this paper proposed to use terminal circuit model and PMU voltage phase angle measurements to simplify power flow calculation. When updating current injections at the right-hand side of power flow equations and using PMU voltage phase angle measurements representing PV buses voltage phase angle, the Jacobian matrix is constant during the iteration. The simplification reduces the computational efforts and improves the computation efficiency. The proposed method is tested on the IEEE 14-bus and IEEE 118-bus standard systems. The results are validated by traditional power flow solution and the computation efficiency is demonstrated.**

## I. INTRODUCTION

Power flow calculation is of great importance as a base case for many other applications, such as contingency analysis, optimal power flow, fault current calculation, and transient stability analysis [1][2][3]. In power system operation practice, generator power output is controlled to meet load demand in megawatts and megavars and to regulate bus voltage magnitude to the target value at PV buses. To get align with the operation practice, the ordinary power flow problem is stated as power balance equation. Since the active power and reactive power are nonlinear to voltages, the power balance equations are nonlinear. Conventionally, Newton-Raphson method and fast decoupled power flow method are adopted to solve the nonlinear power flow equations [4][5].

Power grid is a network composed by resistors, inductors, and capacitors. It is essentially a linear electric circuit subjected to the Kirchhoff's Current Law (KCL) and the Kirchhoff's Voltage Law (KVL). To take the advantages of power network sparsity, current balance equations subjected to KCL usually are used to model power flow problem [6][7]. Formed in the triangular coordinates, the terminal current of power grid components, such as transmission line, transformer, bus shunt is linear to its terminal voltage. This merit can be kept when terminal circuit model and current injections other than power injections are used to describe power flow equations. The nonlinear terms in the current balance equations may be introduced by the components connecting to the grid, such as generator and inconstant impedance load when their current injection is converted from power injection. The nonlinear term in this category can be modeled at right hand side of the current balance equation to keep the Jacobian matrix constant.

The other nonlinear term is introduced at PV bus. The PV bus voltage magnitude is regulated to the target value, thus quadratic term is introduced to the current balance equation. Usually, the nonlinear term cannot be removed from the equation unless both the voltage magnitude and voltage angle at the PV bus are known. Voltage magnitude is given at PV bus which is the target voltage when the regulated generator reactive power output is within limit. However, the PV bus voltage angle is unknown before the power flow problem is solved. In general, generator setup transformer high side bus voltage and pivot bus voltage are regulated, and these buses are PV buses having Phasor Measurement Units (PMU) installed in usual. Assuming the measured voltage phase angle is the solved voltage phase angle, the equations for the PV buses can be removed.

To keep the linear merit of power flow equations to the greatest extent, in this paper, terminal circuit model and current injection-based power flow calculation method is proposed. The current injection is represented at the right-hand side of power flow equations to avoid adding nonlinear terms into the Jacobian matrix. Using the terminal circuit model, the power flow calculation model is modularized and easy to maintain [8]. In terminal circuit model, the calculation of each component is encapsulated in the circuit and interfacing to each other. The interface of the component is the terminal

This work is supported by the State Grid Corporation technology project 5455HJ180021.

voltage and the terminal current. The terminal voltage and the terminal current are subject to the KCL and the KVL. When new components are added to the power system, their model can be created independently and calculated in parallel. If PMUs are fully deployed at PV buses, the Jacobian matrix of current injection-based power flow equations will be constant. This simplified model reuses the Jacobian matrix and its factorization for forward and backward substitution during the iteration to save the computation time significantly.

The remainder of this paper is organized as follows. The terminal circuit model of power network component is discussed in Section II. The current balance equation is introduced in Section III. Current injection of component is described in Section IV. PV bus model is introduced in Section V. Power flow calculation simplification using PMU measurements is discussed in Section VI. The case study results are validated in Section VII. And conclusions are given in Section VIII.

## II. Terminal Circuit Model of Power Network Component

Power network is connected by transmission line, transformer, and bus shunt. The π-type equivalent circuit of transmission line is shown in Figure 1.

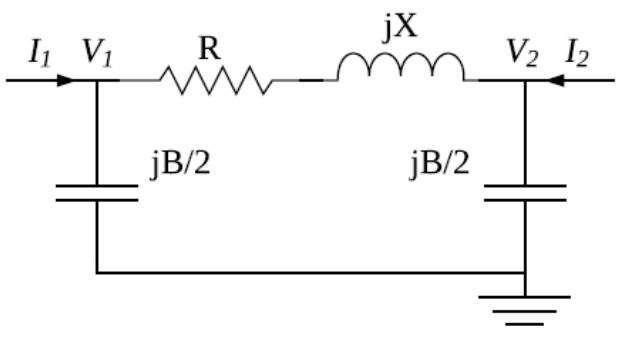

Fig. 1 Transmission Line Equivalent Circuit Model

The equivalent circuit is a linear circuit where the terminal voltage and current are subject to:

$$\begin{bmatrix} I_1 \\ I_2 \end{bmatrix} = \begin{bmatrix} g_{11}+jb_{11} & g_{12}+jb_{12} \\ g_{21}+jb_{21} & g_{22}+jb_{22} \end{bmatrix} \begin{bmatrix} V_1 \\ V_2 \end{bmatrix}$$

$$= \begin{bmatrix} \frac{1}{R+jX}+j\frac{B}{2} & -\frac{1}{R+jX} \\ -\frac{1}{R+jX} & \frac{1}{R+jX}+j\frac{B}{2} \end{bmatrix} \begin{bmatrix} V_1 \\ V_2 \end{bmatrix} \quad (1)$$

The transformer equivalent circuit is shown in Figure 2 and its terminal circuit equation is given by equation (2).

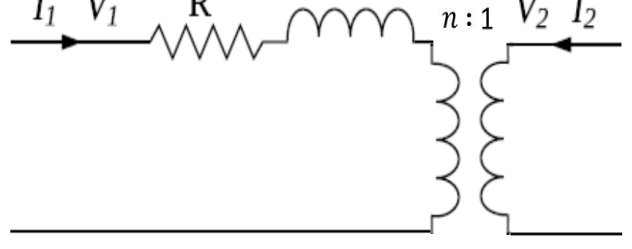

Fig.2 Transformer Equivalent Circuit

$$\begin{bmatrix} I_1 \\ I_2 \end{bmatrix} = \begin{bmatrix} g_{11}+jb_{11} & g_{12}+jb_{12} \\ g_{21}+jb_{21} & g_{22}+jb_{22} \end{bmatrix} \begin{bmatrix} V_1 \\ V_2 \end{bmatrix}$$

$$= \begin{bmatrix} \frac{1}{n^2(R+jX)} & -\frac{1}{n(R+jX)} \\ -\frac{1}{n(R+jX)} & \frac{1}{R+jX} \end{bmatrix} \begin{bmatrix} V_1 \\ V_2 \end{bmatrix} \quad (2)$$

## III. Current Balance Equation

Power system as an electric circuit network is subjected to the Kirchhoff's Current Law. The Kirchhoff's current law specifies that at any node in a circuit, the sum of the branch currents injecting into or absorbing from the node must be equal to zero as expressed by equation (3).

$$YV = I \quad (3)$$

where $YV$ is the summation of branch current. $I$ is the current injections from generator, load, shunt device, and other components at the node. The equation can be rewritten in matrix format by equation (4) and in rectangular format by equation (5).

$$\begin{bmatrix} G_{11}+jB_{11} & G_{12}+jB_{12} & \cdots & G_{1n}+jG_{1n} \\ G_{21}+jG_{21} & G_{22}+jB_{22} & \cdots & G_{2n}+jB_{2n} \\ \vdots & \vdots & \ddots & \vdots \\ G_{n1}+jB_{n1} & G_{n2}+jB_{n2} & \cdots & G_{nn}+jB_{nn} \end{bmatrix} \begin{bmatrix} V_{1x}+jV_{1y} \\ V_{2x}+jV_{2y} \\ \vdots \\ V_{nx}+jV_{ny} \end{bmatrix} = \begin{bmatrix} I_{1x}+jI_{1y} \\ I_{2x}+jI_{2y} \\ \vdots \\ I_{nx}+jI_{ny} \end{bmatrix} \quad (4)$$

$$\begin{bmatrix} \begin{bmatrix} G_{11} & -B_{11} \\ B_{11} & G_{11} \end{bmatrix} & \begin{bmatrix} G_{12} & -B_{12} \\ B_{12} & G_{12} \end{bmatrix} & \cdots & \begin{bmatrix} G_{1n} & -B_{1n} \\ B_{1n} & G_{1n} \end{bmatrix} \\ \begin{bmatrix} G_{21} & -B_{21} \\ B_{21} & G_{21} \end{bmatrix} & \begin{bmatrix} G_{22} & -B_{22} \\ B_{22} & G_{22} \end{bmatrix} & \cdots & \begin{bmatrix} G_{2n} & -B_{2n} \\ B_{2n} & G_{2n} \end{bmatrix} \\ \vdots & \vdots & \ddots & \vdots \\ \begin{bmatrix} G_{n1} & -B_{n1} \\ B_{n1} & G_{n1} \end{bmatrix} & \begin{bmatrix} G_{n2} & -B_{n2} \\ B_{n2} & G_{n2} \end{bmatrix} & \cdots & \begin{bmatrix} G_{nn} & -B_{nn} \\ B_{nn} & G_{nn} \end{bmatrix} \end{bmatrix} \begin{bmatrix} \begin{bmatrix} V_{1x} \\ V_{1y} \end{bmatrix} \\ \begin{bmatrix} V_{2x} \\ V_{2y} \end{bmatrix} \\ \vdots \\ \begin{bmatrix} V_{nx} \\ V_{ny} \end{bmatrix} \end{bmatrix} = \begin{bmatrix} \begin{bmatrix} I_{1x} \\ I_{1y} \end{bmatrix} \\ \begin{bmatrix} I_{2x} \\ I_{2y} \end{bmatrix} \\ \vdots \\ \begin{bmatrix} I_{nx} \\ I_{ny} \end{bmatrix} \end{bmatrix} \quad (5)$$

where the non-diagonal element $G_{ij}+jB_{ij}$ is the admittance of branch as derived in equation (1) for transmission line and in equation (2) for transformer with the turn ratio of transformer represented by $n$.

The diagonal element $G_{ii}+jB_{ii}$ is the summation of the non-diagonal elements and the bus shunt $B_{sh}$ at the bus in equation (6).

$$\begin{bmatrix} G_{ii} & -B_{ii} \\ B_{ii} & G_{ii} \end{bmatrix} = \begin{bmatrix} -\sum G_{ij} & \sum B_{ij} - B_{sh,i} \\ -\sum B_{ij} + B_{sh,i} & -\sum G_{ij} \end{bmatrix} \quad (6)$$

The Y matrix is constant, but the current injections at right hand side of equation (5) may be nonlinear to the terminal voltage when the load is not the constant impedance load. To solve the nonlinear equations, Newton method is usually adopted to update the voltage by using the update equation (7). The updated voltages are used in return to calculate the current injection either at right hand side or in the Jacobian matrix till the solution is achieved.

$$Y\Delta V = \Delta I \quad (7)$$

## IV. Current Injection of Component

In power systems, loads, generators, and bus shunts contribute current injections to buses. Bus shunt $B_{sh}$ current injection is:

$$\begin{bmatrix} I_x \\ I_y \end{bmatrix} = \begin{bmatrix} B_{sh} & 0 \\ 0 & -B_{sh} \end{bmatrix} \begin{bmatrix} V_x \\ V_y \end{bmatrix} \tag{8}$$

The current in equation (8) has been represented by $B_{sh}$ in (6) already. The shunt current contribution does not need to duplicate to the right-hand side of equation (5).

Generator and load current injections are calculated by equation (9) and equation (10) separately and summed together by equation (11) assuming the generation output is $P_g + jQ_g$, the load demand is $P_l + jQ_l$, and the bus voltage is $V_x + jV_y$.

$$\begin{bmatrix} I_{gx} \\ I_{gy} \end{bmatrix} = \begin{bmatrix} \frac{P_g}{V_x^2+V_y^2} & \frac{Q_g}{V_x^2+V_y^2} \\ \frac{-Q_g}{V_x^2+V_y^2} & \frac{P_g}{V_x^2+V_y^2} \end{bmatrix} \begin{bmatrix} V_x \\ V_y \end{bmatrix} \tag{9}$$

$$\begin{bmatrix} I_{lx} \\ I_{ly} \end{bmatrix} = \begin{bmatrix} \frac{-P_l}{V_x^2+V_y^2} & \frac{-Q_l}{V_x^2+V_y^2} \\ \frac{Q_l}{V_x^2+V_y^2} & \frac{-P_l}{V_x^2+V_y^2} \end{bmatrix} \begin{bmatrix} V_x \\ V_y \end{bmatrix} \tag{10}$$

$$\begin{bmatrix} I_x \\ I_y \end{bmatrix} = \begin{bmatrix} I_{gx} - I_{lx} \\ I_{gy} - I_{ly} \end{bmatrix} = \begin{bmatrix} \frac{P_s}{V_x^2+V_y^2} & \frac{Q_s}{V_x^2+V_y^2} \\ \frac{-Q_{is}}{V_x^2+V_y^2} & \frac{P_{is}}{V_x^2+V_y^2} \end{bmatrix} \begin{bmatrix} V_x \\ V_y \end{bmatrix} \tag{11}$$

$$P_s + jQ_s = (P_g - P_l) + j(Q_g - Q_l) \tag{12}$$

Obviously, the generator and load current injections are nonlinear to the terminal voltage. Taking account of the current injection impacts, the voltage update equation is revised as:

$$\left[ Y - \frac{\partial I(V)}{\partial V} \right] \Delta V = I(V) - YV \tag{13}$$

where,

$$\left[ \frac{\partial I(V)}{\partial V} \right] = \begin{bmatrix} \frac{\partial I_x(V)}{\partial V_x} & \frac{\partial I_x(V)}{\partial V_y} \\ \frac{\partial I_y(V)}{\partial V_x} & \frac{\partial I_y(V)}{\partial V_y} \end{bmatrix} = \begin{bmatrix} \frac{P_s(V_x^2-V_y^2)-2Q_sV_xV_y}{(V_x^2+V_y^2)^2} & \frac{Q_s(V_x^2-V_y^2)-2P_sV_xV_y}{(V_x^2+V_y^2)^2} \\ \frac{Q_s(V_x^2-V_y^2)-2P_sV_xV_y}{(V_x^2+V_y^2)^2} & -\frac{P_s(V_x^2-V_y^2)-2Q_sV_xV_y}{(V_x^2+V_y^2)^2} \end{bmatrix} \tag{14}$$

The current injections in (11) can be represented in two ways in power flow equation (5), i.e. the linear representation and the nonlinear representation. In the linear representation, the nonlinear current injections are represented at the right-hand side of the equation (5) to keep the Jacobian matrix constant. In the nonlinear representation, the nonlinear terms (14) are introduced to the Jacobian matrix in (13) to make the Jacobian matrix a nonlinear function of bus voltage. The linear representation requires less computation in each iteration since the Jacobian matrix is constant and the Jacobian matrix refactorization is unnecessary. The nonlinear representation will update the Jacobian matrix and redo the factorization in each iteration. Power flow calculation has linear convergence using the linear representation and quadratic convergence using the nonlinear representation. Power flow calculation may need more iterations using the linear representation. The computation performance of the two representations will be compared in the case study section.

## V. PV Bus Model

At the PV bus, the voltage magnitude is regulated to the target voltage $V_t$.

$$V_x = \sqrt{V_t^2 - V_y^2} \tag{15}$$

Substitute equation (15) to equation (11), we get

$$\begin{bmatrix} I_x \\ I_y \end{bmatrix} = \begin{bmatrix} \frac{P_s\sqrt{V_t^2 - V_y^2} + Q_sV_y}{V_t^2} \\ \frac{-Q_s\sqrt{V_t^2 - V_y^2} + P_sV_y}{V_t^2} \end{bmatrix} \tag{16}$$

In equation (16), the injection current is a function of $V_y$ and $Q_s$, and $Q_s$ is the unknow variable at PV bus. The $V_y$ and $Q_s$ update equations are derived as:

$$\begin{bmatrix} \Delta I_x \\ \Delta I_y \end{bmatrix} = \begin{bmatrix} \frac{Q_s\frac{V_y}{V_x} + P_s}{V_t^2} & \frac{V_x}{V_t^2} \\ \frac{P_s\frac{V_y}{V_x} + Q_s}{V_t^2} & \frac{V_y}{V_t^2} \end{bmatrix} \begin{bmatrix} \Delta Q_s \\ \Delta V_y \end{bmatrix} = \begin{bmatrix} a & b \\ c & d \end{bmatrix} \begin{bmatrix} \Delta Q_s \\ \Delta V_y \end{bmatrix} \tag{17}$$

To generalize, assume the bus $k$ is a PV bus. The voltage updated equation (7) is changed to equation (18).

$$\begin{bmatrix} \begin{bmatrix} G_{11} & -B_{11} \\ B_{11} & G_{11} \end{bmatrix} & \cdots & \begin{bmatrix} a_k & -B_{1k}+b_k \\ c_k & G_{1k}+d_k \end{bmatrix} & \cdots & \begin{bmatrix} G_{1n} & -B_{1n} \\ B_{1n} & G_{1n} \end{bmatrix} \\ \vdots & \ddots & \vdots & \ddots & \vdots \\ \begin{bmatrix} G_{k1} & -B_{k1} \\ B_{k1} & G_{k1} \end{bmatrix} & \cdots & \begin{bmatrix} a_k & -B_{kk}+b_k \\ c_k & G_{kk}+d_k \end{bmatrix} & \cdots & \begin{bmatrix} G_{kn} & -B_{kn} \\ B_{kn} & G_{kn} \end{bmatrix} \\ \vdots & \ddots & \vdots & \ddots & \vdots \\ \begin{bmatrix} G_{n1} & -B_{n1} \\ B_{n1} & G_{n1} \end{bmatrix} & \cdots & \begin{bmatrix} a_k & -B_{nk}+b_k \\ c_k & G_{nk}+d_k \end{bmatrix} & \cdots & \begin{bmatrix} G_{nn} & -B_{nn} \\ B_{nn} & G_{nn} \end{bmatrix} \end{bmatrix} \begin{bmatrix} \begin{bmatrix} \Delta V_{1x} \\ \Delta V_{1y} \end{bmatrix} \\ \vdots \\ \begin{bmatrix} \Delta Q_{ks} \\ \Delta V_{ky} \end{bmatrix} \\ \vdots \\ \begin{bmatrix} \Delta V_{nx} \\ \Delta V_{ny} \end{bmatrix} \end{bmatrix} = \begin{bmatrix} \begin{bmatrix} \Delta I_{1x} \\ \Delta I_{1y} \end{bmatrix} \\ \vdots \\ \begin{bmatrix} \Delta I_{kx} \\ \Delta I_{ky} \end{bmatrix} \\ \vdots \\ \begin{bmatrix} \Delta I_{kx} \\ \Delta I_{ky} \end{bmatrix} \end{bmatrix} \tag{18}$$

The four new terms $a_k$, $b_k$, $c_k$, and $d_k$ in equation (18) are non-constant. They are function of the bus voltage and reactive power injection of generator at regulating bus. They are calculated in each iteration till the voltage is converged. Once $V_{ky}$ at PV bus is updated, $V_{kx}$ can be calculated by (15) in each iteration.

## VI. Simplify Power Flow Calculation by PMU Measurements

The Jacobian matrix in equation (18) is non-constant when the four new terms are introduced to represent PV bus. When taking the target voltage as PV bus voltage magnitude (PV bus will be switched to PQ bus when the generator reactive power limit is reached), and taking the PMU voltage phase angle measurement as PV bus voltage angle, the PV bus voltage is a given. Thus, the columns and rows in the equation for the PV bus can be removed. The power flow equation (18) is then simplified as

$$\begin{bmatrix} \begin{bmatrix} G_{11} & -B_{11} \\ B_{11} & G_{11} \end{bmatrix} & \cdots & \cdots & \cdots & \begin{bmatrix} G_{1n} & -B_{1n} \\ B_{1n} & G_{1n} \end{bmatrix} \\ \vdots & \ddots & \vdots & \ddots & \vdots \\ \vdots & \cdots & \vdots & \cdots & \vdots \\ \vdots & \ddots & \vdots & \ddots & \vdots \\ \begin{bmatrix} G_{n1} & -B_{n1} \\ B_{n1} & G_{n1} \end{bmatrix} & \cdots & \cdots & \cdots & \begin{bmatrix} G_{nn} & -B_{nn} \\ B_{nn} & G_{nn} \end{bmatrix} \end{bmatrix} \begin{bmatrix} \begin{bmatrix} \Delta V_{1x} \\ \Delta V_{1y} \end{bmatrix} \\ \vdots \\ \vdots \\ \vdots \\ \begin{bmatrix} \Delta V_{nx} \\ \Delta V_{ny} \end{bmatrix} \end{bmatrix} = \begin{bmatrix} \begin{bmatrix} \Delta I_{1x} \\ \Delta I_{1y} \end{bmatrix} \\ \vdots \\ \vdots \\ \vdots \\ \begin{bmatrix} \Delta I_{kx} \\ \Delta I_{ky} \end{bmatrix} \end{bmatrix} \quad (19)$$

Using PMU measurements along with the linear representation of the generator and the load current injections, the Jacobian matrix of the power flow equation (19) will be constant which does not need to be updated during the iterations and the power flow calculation is greatly simplified.

## VII. NUMERICAL CASE STUDY

The standard IEEE 14-bus system and the IEEE 118-bus systems are employed to demonstrate the computation efficiency and accuracy of the proposed power flow calculation method. The testing environment is as listed in Table I.

TABLE I. HARDWARE AND SOFTWARE TESTING ENVIRONMENT

| | |
|---|---|
| **Software Environment** | |
| Operation System | Windows 7 Professional |
| **Hardware Environment** | |
| CPU | Intel® Core i7-6600U @2.60 GHz |
| Memory | 16GB |

In order to validate the proposed method, a prototype computer program for the proposed method and the conventional power flow calculation method are developed. To make the two methods comparable, the Jacobian matrix formation, LU decomposition and forward/backward substitution are developed as common functions to be called by both methods. The difference is the proposed method calls the Jacobian matrix formation and the LU decomposition just once, then keeps them as the same for all iterations. But the conventional power flow calculation needs to update the Jacobian matrix, and the LU decomposition in each iteration. Using the two different methods, the study results of the proposed method are compared against the conventional power flow calculation method and are shown in the Table II.

TABLE II. COMPUTATION EFFICIENCY COMPARISON

| System | Method | Power Convergence Tolerance (p.u.) | Number of Iterations | Computation Time (ms) |
|---|---|---|---|---|
| IEEE14 | Traditional Method | 1.00E-05 | 4 | 21.3 |
| | Proposed Method | 1.00E-05 | 5 | 10.2 |
| IEEE118 | Traditional Method | 1.00E-05 | 4 | 168.5 |
| | Proposed Method | 1.00E-05 | 5 | 72.2 |

Shown in the Table II, by the proposed method, power flow calculation needs one more iteration to converge for IEEE 14-bus system because the current injection impacts are represented at right hand side of the power flow equations. However, since the proposed method use the constant Jacobian matrix, and does not require to refactorize the Jacobian Matrix which results in less computation time. Comparing with the traditional method, the proposed method saves more than 50% of the total computation time. The execution time is separated in Table III for further investigation.

TABLE III. SEPARATED EXECUTION TIME (ms)

| System | Method | Jacobian Matrix Formation | LU Decomposition | F/B Substitution |
|---|---|---|---|---|
| IEEE14 | Traditional Method | 3.9 | 13.8 | 1.4 |
| | Proposed Method | 1.4 | 3.4 | 3.3 |
| IEEE118 | Traditional Method | 13.2 | 146.1 | 6.6 |
| | Proposed Method | 2.8 | 49.2 | 9.1 |

In the traditional method, the diagonal elements of the Jacobian matrix need to be updated in each iteration, so as to the LU decomposition. The total CPU time for the Jacobian matrix formation and LU decomposition is higher than that by the proposed method since proposed method keeps the Jacobian matrix constant. The traditional method takes 4 iterations for the IEEE 14 bus system and the IEEE 118 bus system to converge the power flow calculation. In contrary, the proposed method takes one more iteration to converge. The CPU time of the forward and backward substitution for the proposed method is longer than that of the traditional method. But the total computation time is compensated by the saving on the Jacobian matrix formation and the LU decomposition.

To compare the convergence of the two methods, the maximum voltage difference in each iteration by the two methods for the IEEE 14-bus system and the IEEE 118-bus system are illustrated in the Figure 3 and Figure 4. The traditional method shows better convergence.

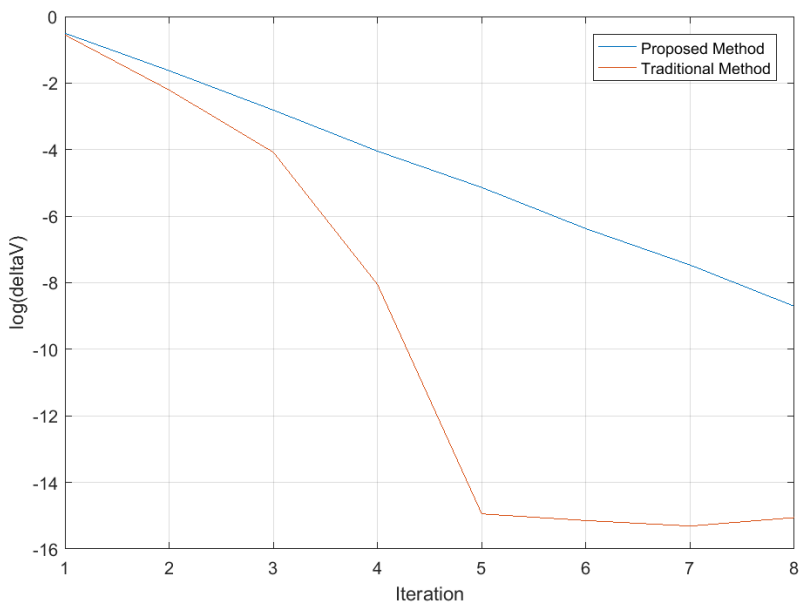

Fig.3 Power Flow Convergence for IEEE 14 Bus System

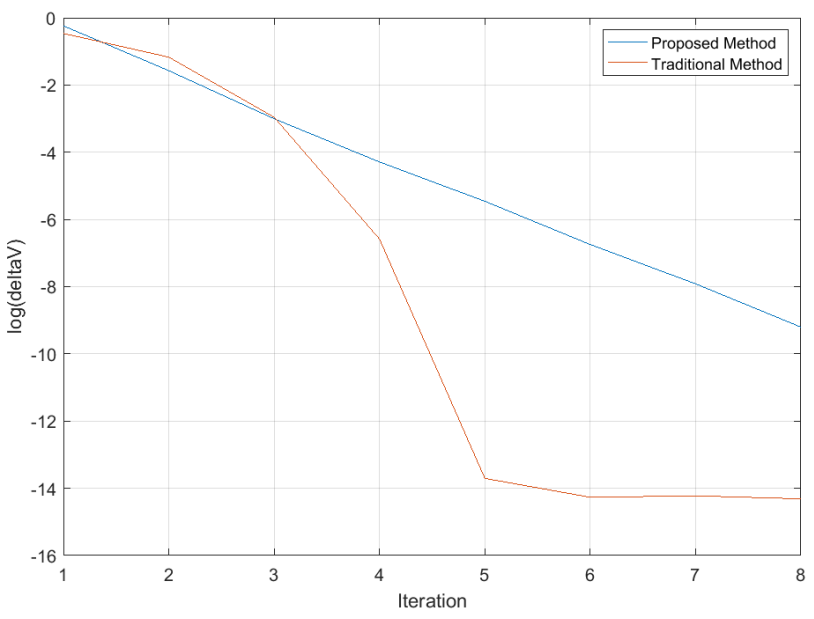

Fig.4 Power Flow Convergence for IEEE 118 Bus System

The proposed method assumes the PV bus voltage phase angle is equal to the corresponding PMU voltage phase angle measurement. However, PMU has measurement error. According to the IEEE standard C37.118.1, the steady-state compliance should be confirmed that the phase angle measurement maximum total vector error (TVE) is 1% [9]. In the case study, the PMU measurement is simulated by adding TVE to the phase angle solution by the conventional power flow calculation. To simulate the impact of the PMU voltage phase angle measurement error, the worst scenario and the random scenario are created. In the worst scenario, all PMU voltage phase angle measurement has the maximum TVE (±1%). In the random scenario, a random TVE meeting unit distribution (-1%, 1%) is applied. Under the two scenarios, the voltage maximum absolute error of the proposed method to the conventional power flow calculated method for the two studied systems are shown in the Table IV.

TABLE IV. COMPUTATION ACCURACY

| System | Scenario | Voltage Magnitude Maximum Absolute Error (p.u.) | Voltage Angle Maximum Absolute Error (rad) |
|---|---|---|---|
| IEEE14 | Worst Scenario | 0.00029 | 0.00267 |
| | Random Scenario | 0.00006 | 0.00096 |
| IEEE118 | Worst Scenario | 0.00047 | 0.00637 |
| | Random Scenario | 0.00030 | 0.00240 |

The maximum absolute errors for voltage magnitude and voltage angle are less than 0.0005 p.u. and 0.007 rad which are satisfied for operation. The detailed absolute errors for the IEEE 118-bus system in the two scenarios at each bus are shown in the Figure 5. The averaged absolute errors for the two cases in the two scenarios are less than 0.0001 p.u. and 0.003 rad respectively.

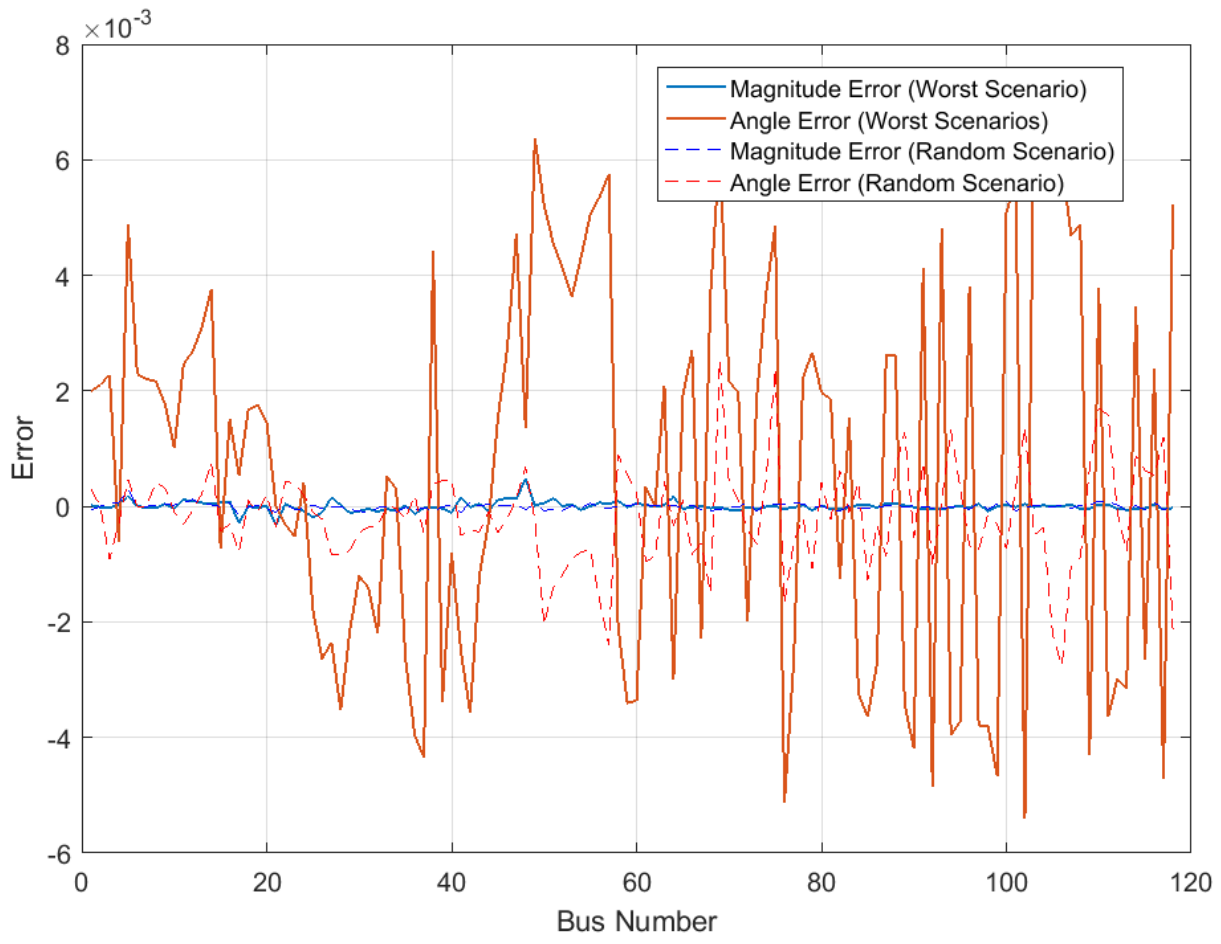

Fig.5 Voltage Error for IEEE 118 Bus System

## VIII. CONCLUSION

The performance of power flow calculation is critical in real time application. To improve the computation efficiency of power flow calculation without compromising the calculation accuracy, current injection-based power flow calculation method is proposed with two features to keep the Jacobian matrix constant. The nonlinear current injections are calculated at right hand side of the equation. And the PMU voltage phase angle measurements are applied to represent PV bus voltage angle. The second feature may not be always true but taking PMU voltage phase angle measurement account will simplify the power flow calculation to the greatest extend. The current injection-based network solution method is a good fit for transient stability analysis when sequential method is adopted since the algebraic equations representing the network solution and the differential equations representing the transient behavior of dynamic components are solved alternatively in the sequential method. The method proposed in this paper will save the computation time for solving the algebraic equations.

The case study validates the efficiency of the proposed method. The proposed method is outperformed than the traditional method in terms of computation efficiency. The computation time is reduced by the proposed method to half and the maximum absolute error in the worst case is less than 0.0005 p.u. for voltage magnitude and 0.007 rad for voltage phase angle. The proposed method is promising to be applied to real-time power flow calculation and on-line transient stability analysis.